\documentclass[preprint,showpacs,preprintnumbers,amsmath,amssymb]{revtex4}
\usepackage{epsfig}
\usepackage{graphicx} 
\usepackage{bm}

\begin{document}

\title{Laser scattering by density fluctuations of ultra-cold atoms in a magneto-optical trap}

\author{J.T. Mendon\c{c}a$^1$}
\email{titomend@ist.utl.pt}
\author{H. Ter\c{c}as$^2$}

\affiliation{IPFN$^1$ and CFIF$^2$, Instituto Superior T\'ecnico, 1049-001 Lisboa, Portugal}


\begin{abstract}

We study the spectrum of density fluctuations in the ultra-cold gas of neutral atoms, confined in a magneto-optical trap. We determine the corresponding amplitude and spectra of laser light scattered by this medium. We derive an expression for the dynamical structure function, by using a test particle method. We propose to use the collective laser scattering as a diagnostic method for the microscopic properties of the ultra-cold matter. This will also allow us to check on the atomic correlations which are mediated by the collective mean field inside the gas.

\end{abstract}
\pacs{}

\maketitle

\section{Introduction}

In recent years, attention has been given to the study of collective oscilations in alkali metal gases, cooled and confined  in magneto-optical traps (MOTs) \cite{kaiser,kaiser2,distefano}. Such a cooperative behavior is driven by light and results from the random scattering between the atoms, being in the root of a long-range interaction potential. Surprisingly, this collective interaction is very similar to that observed in Coulomb systems, since it scales as $1/r^2$, which allows one to establish analogies between cold atoms and plasmas. 

The evidence of such Coulomb-like forces due the random scattering of photons was first discovered by Walker et al. \cite{walker} and Dalibard \cite{dalibard}, where the divergence of the effective net force between the atoms was shown to be proportional to the MOT density. As a consequence, it is possible to define an equivalent electric charge for the neutral atoms. The advantage of such a system is that the effective charge depends on the basic parameters of the system, what makes MOTs good candidates to observe plasma physics phenomena with a tunable ''electric'' charge. A strong manifestation of the plasma character of MOTs was also revealed by the studies of Coulomb explosions of the atomic clouds,  performed by Pruvost et al \cite{pruvost}. Recently, we have shown theoretically that waves and oscillations of the plasma type can also exist in the neutral gas \cite{mendkaiser,mend2010} and have explored the plasma character of magneto-optical traps to the study of parametric instabilities \cite{hugo}.

Here we propose to study the spectrum of density fluctuations in the ultra-cold gas, by using collective laser scattering techniques. In doing this, we explore even further the analogy between the ultra-cold neutral gas confined in a magneto-optical trap and a plasma, by adapting well known plasma physics theoretical techniques to this new problem. First, we determine the scattered fields, and the expected average scattered power. Second, we determine the spectrum of density fluctuations, in the quasi-classical approximation. We use here the test particle method, well known in plasma physics \cite{book}. Finally, we characterize the main properties of the expected scattered signal.

\section{Scattered laser power}

We start by stating the basic equations of our problem. The propagation of a laser pulse along the ultra-cold gas is governed by the wave equation

\begin{equation}
\left( \nabla^2 - \frac{1}{c^2} \frac{\partial^2}{\partial t^2} \right) \mathbf{E} = - \mu_0 \frac{\partial^2 \mathbf{P}}{\partial t^2}
\label{eq:2.1} \end{equation}
where $\mathbf{E}$ is the laser electric field and $\mathbf{P}$ is the polarization field in the gas, as determined by

\begin{equation}
\mathbf{P} (\mathbf{r}, t)  = \epsilon_0 \int_0^\infty \bm{\Pi} (\mathbf{r}, \tau)\cdot \mathbf{E} (\mathbf{r}, t - \tau) d \tau,
\label{eq:2.1b} 
\end{equation}
where $\bm {\Pi} (\mathbf{r}, t)$ is the atomic susceptibility tensor. Using Fourier transformations of the fields

\begin{equation}
\mathbf{E} (\mathbf{r}, t)  = \int \mathbf{E} (\mathbf{r}, \omega) \exp ( - i \omega t) \; \frac{d \omega}{2 \pi},
\label{eq:2.2} 
\end{equation}
one obtains

\begin{equation}
\mathbf{P} (\mathbf{r}, \omega)  = \epsilon_0 \bm{\Pi} (\mathbf{r}, \omega)\cdot \mathbf{E} (\mathbf{r}, \omega).
\label{eq:2.3} 
\end{equation}
Because we can neglect the anisotropies in the majority of the MOT, the atomic susceptibility tensor $\bm{\Pi}$ can be safely replaced  by the spectral susceptibility function $\chi$ which, for the two-level atom, is given by

\begin{equation}
\chi (\mathbf{r}, \omega)  = n (\mathbf{r}) \chi_a (\omega) \; , \quad \chi_a (\omega) = \frac{\Omega^2}{3 \epsilon_0 \hbar} \frac{\delta + i \Gamma}{\delta^2 + \Gamma^2}
\label{eq:2.3b} 
\end{equation}
Here, $n (\mathbf{r})$ is the number of atoms per unit volume, $\chi_a (\omega)$ is the single atom susceptibility, $\delta \equiv \omega - \omega_{a}$ is the radiation frequency detuning with respect to the atomic transition frequency $\omega_{a}$, $\Gamma$ is the corresponding line width and $\Omega$ represents the Rabi frequency. The local density can fluctuate in time, at frequencies much lower than those of the radiation field frequency $\omega$, thus leading to the replacement of $n (\mathbf{r})$ by $n (\mathbf{r}, t)$ in the above expression. The total radiation field $\mathbf{E}$ will be determined by the sum of two parts, the incident laser field $\mathbf{E}_0$, and the scattered field $\mathbf{E}_s$. For a real incident laser field, we can use

\begin{equation}
\mathbf{E} (\mathbf{r}, t)  = \frac{1}{2} \mathbf{E}_0 \exp (i \mathbf{k}_0 \cdot \mathbf{r} - i \omega_0 t ) + c.c.,
\label{eq:2.4} \end{equation}
where the wave-vector ${\bf k}_0$ follows the dispersion relation

\begin{equation}
k_0^2 = \frac{\omega_0^2}{c^2} \left[ 1 + \chi (\omega_0) \right] \; , \quad \chi (\omega_0) = n_0 \chi_a (\omega_0),
\label{eq:2.4b} 
\end{equation}
with $n_0$ being the atomic mean density. Such an approximation is valid for MOTs with a moderate number of atoms (typically, $N\sim 10^{8}$.) On the other hand, the scattered field can be generically written as

\begin{equation}
\mathbf{E}_s (\mathbf{r}, t)  = \int \mathbf{E}_s (\mathbf{r}, \omega)  \exp (- i \omega t )  \frac{d \omega}{2 \pi}
\label{eq:2.5} 
\end{equation}
and the field components can be determined by solving the corresponding wave equation, which can be written as

\begin{equation}
\left( \nabla^2 + k^2 \right) \mathbf{E}_s (\mathbf{r}, \omega) = \frac{\omega^2}{2 c^2} \mathbf{E}_0 e^{i \mathbf{k}_0 \cdot \mathbf{r}} n (\mathbf{r}, \omega') + c.c.,
\label{eq:2.6} 
\end{equation}
 where

\begin{equation}
n (\mathbf{r}, \omega') = \int n (\mathbf{r}, t ) e^{i \omega' t} d t
\label{eq:2.6b} 
\end{equation}
and $\omega' = \omega - \omega_0$ is the frequency of density fluctuation. In its turn, the scattered wave number $\mathbf{k}$ will be determined by the same dispersion relation as in Eq. (\ref{eq:2.4b}), with $k_0$ and $\omega_0$ respectively replaced  by $k$ and $\omega$. The solution to Eq. (\ref{eq:2.6}) can therefore be written as

\begin{equation}
E_s (\mathbf{r}, \omega) = - \frac{i \omega^2}{2 k c^2} \chi_a (\omega_0) (\mathbf{e}_\omega \cdot \mathbf{e}_0) E_0 e^{i \mathbf{k} \cdot \mathbf{r}} \int d \mathbf{r} \int \frac{d \mathbf{k}'}{(2 \pi)^3} n ( \mathbf{k}', \omega') e^{ i (\mathbf{k}_0 + \mathbf{k}' - \mathbf{k}) \cdot \mathbf{r}} + c.c.,
\label{eq:2.7} 
\end{equation}
where we have used the unit polarization vectors for the incident and the scattered fields, $\mathbf{e}_0 = \mathbf{E}_0 / E_0$ and $\mathbf{e}_\omega = \mathbf{E}_s (\omega) / E_s (\omega)$. The integration in volume leads to the appearance of a delta function $\delta (\mathbf{k}_0 + \mathbf{k}' - \mathbf{k})$, and subsequent integration over $\mathbf{k}'$ finally leads to

\begin{equation}
E_s (\mathbf{r}, \omega) = - i A_s \exp (i \mathbf{k} \cdot \mathbf{r}) + c.c.,
\label{eq:2.8} \end{equation}
where the amplitude of the scattered field is determined by

\begin{equation}
A_s ( \omega) = \frac{\omega^2}{2 k c^2} \chi_a (\omega_0) (\mathbf{e}_\omega \cdot \mathbf{e}_0) E_0 n (\omega', \mathbf{k}')
\label{eq:2.9} 
\end{equation}
with $\omega = \omega_0 + \omega'$ and $\mathbf{k} = \mathbf{k}_0 + \mathbf{k}'$. The relative scattered intensity along the direction of $\mathbf{k}$ is therefore determined by the dynamical structure factor

\begin{equation}
S(\mathbf{k},\omega)=\frac{| E_s (\mathbf{k}, \omega) |^2}{| E_0 |^2} =  \frac{i \omega^4}{4 k^2 c^4} | \chi_a (\omega_0) |^2 (\mathbf{e}_\omega \cdot \mathbf{e}_0)^2  | n ( \mathbf{k}', \omega') |^2. 
\label{eq:2.10}
 \end{equation}
This quantity can easily be related with the density correlation function, by noting that

\begin{equation}
| n (\mathbf{k}', \omega') |^2 = \int d \mathbf{r}_1 d t_1 \int d \mathbf{r}_2 d t_2 n (\mathbf{r}_1, t_1) e^{- i \mathbf{k}' \cdot \mathbf{r}_1 + i \omega' t_1} \; n (\mathbf{r}_2, t_2) e^{i \mathbf{k}' \cdot \mathbf{r}_2 - i \omega' t_2}. 
\label{eq:2.11a} 
\end{equation}
Performing an average in both space and time, the latter equation reads

\begin{equation}
< | n (\mathbf{k}', \omega') |^2 > = \lim_{V, T \rightarrow \infty} \frac{1}{V T} \int d \mathbf{r}_1 d t_1 \int d \mathbf{r}_2 d t_2 < n (\mathbf{r}_1, t_1)  n (\mathbf{r}_2, t_2 ) > e^{i \mathbf{k}' \cdot (\mathbf{r}_2 - \mathbf{r}_1)} e^{ - i \omega' (t_2 - t_1)} 
\label{eq:2.11b} 
\end{equation}
Equation (\ref{eq:2.10}) establishes the relation between the scattered signal and the fluctuations inside the cloud, which contain the basic features of the laser cooled gas. We should therefore proceed by understanding how to compute such fluctuations by taking into account the microscopic distribution of the atomic states. 

\section{Test particle method}

In the quasi-classical approximation, a trapped alkali gas in a MOT can be described by a kinetic equation of the form \cite{mendkaiser}

\begin{equation}
\left( \frac{\partial}{\partial t} + \mathbf{v} \cdot \nabla + \frac{\mathbf{F}}{M} \cdot \frac{\partial}{\partial \mathbf{v}} \right) W = 0,
\label{eq:3.1} 
\end{equation}
where $W \equiv W (\mathbf{r}, \mathbf{v}, t)$ is the atomic distribution function. Here the collective force $\mathbf{F}$ can be determined by the Poisson equation
\begin{equation}
\nabla \cdot \mathbf{F} = Q n \equiv Q \int W (\mathbf{r}, \mathbf{v}, t) d \mathbf{v},
\label{eq:3.2} \end{equation}
where $Q$ is the atomic effective charge \cite{mendkaiser, pruvost}

\begin{equation}
Q=\frac{I_{0}}{c}\sigma_{R}(\sigma_{R}-\sigma_{L}).
\label{eq:3.3}
\end{equation}
Here, $I_{0}=2I_{sat}\Omega^2/\Gamma^2$ is the total intensity resulting from the six cooling laser beams, $I_{sat}$ is the saturation intensity of the atom, $c$ is the speed of light, $\sigma_{L}$ represents the optical cross section for absorbing photons of the laser beams and $\sigma_{R}$ is the optical cross section for absorbing photons that are reradiated from the atoms in the trap. \cite{towsend}. Our procedure will consist in the test particle method \cite{book}, by considering a single atom inside the trap moving with speed $\mathbf{v}'$ as the perturbation source. The presence of such a test particle will lead to a fluctuation in the gas distribution function $\tilde{W}$, as determined by the linearized kinetic equation

\begin{equation}
\left( \frac{\partial}{\partial t} + \mathbf{v} \cdot \nabla \right) \tilde{W} = - \frac{\delta \mathbf{F}}{M} \cdot \frac{\partial}{\partial \mathbf{v}}  W_0,
\label{eq:3.3} \end{equation}
where $W_0 = W - \tilde{W}$ is the unperturbed (or the equilibrium) distribution function, and the collective force perturbation $\delta \mathbf{F}$ induced by the test particle in the medium is determined by 

\begin{equation}
\nabla \cdot \delta \mathbf{F} = Q  \int \left[  \tilde{W} + \delta (\mathbf{r} - \mathbf{r}') \delta (\mathbf{v} - \mathbf{v}') \right] d \mathbf{v}.
\label{eq:3.4} \end{equation}
The test particle position is assumed to be given by $\mathbf{r}' \equiv \mathbf{r}' (t) = \mathbf{r}_0 + \mathbf{v}' t$. Let us multiply equation (\ref{eq:3.4}) by $\exp (- i \mathbf{k} \cdot \mathbf{r} + i \omega t)$ and perfomr integrations over both space and time. Defining the Fourier components $\delta \mathbf{F}_{\omega \mathbf{k}}$ and  $\tilde{W}_{\omega \mathbf{k}}$ for the perturbed force and distribution, respectively, we can easily obtain

\begin{equation}
i \mathbf{k} \cdot \delta \mathbf{F}_{\omega \mathbf{k}} = 2 \pi Q \delta (\omega - \mathbf{k} \cdot \mathbf{v}') + Q \int \tilde{W}_{\omega \mathbf{k}} (\mathbf{v} d \mathbf{v}
\label{eq:3.6} 
\end{equation}

Similarly, the Fourier transformation of equation (\ref{eq:3.3}) leads to
\begin{equation}
\tilde{W}_{\omega \mathbf{k}} =  - \frac{i}{M} \frac{\delta \mathbf{F}_{\omega \mathbf{k}} \cdot \partial W_0 / \partial \mathbf{v}}{(\omega - \mathbf{k} \cdot \mathbf{v})}
\label{eq:3.7} 
\end{equation}
Replacing this in equation (\ref{eq:3.6}), and noting that the collective force is purely longitudinal, in such a way that $\delta \mathbf{F}_{\omega \mathbf{k}} = \mathbf{k} \delta F_{\omega \mathbf{k}} / k$, we can write 

\begin{equation}
\delta \mathbf{F}_{\omega \mathbf{k}} = - i \frac{2 \pi Q}{k^2}  \frac{\delta (\omega - \mathbf{k} \cdot \mathbf{v}')}{\epsilon (\omega, \mathbf{k})} \mathbf{k},
\label{eq:3.8}
 \end{equation}
where we have introduced the dielectric function of the atomic gas 

\begin{equation}
\epsilon (\omega, \mathbf{k}) = 1+\chi(\omega,\mathbf{k})=1 + \frac{Q}{M k^2} \int \frac{\mathbf{k} \cdot \partial W_0 / \partial \mathbf{v}}{(\omega - \mathbf{k} \cdot \mathbf{v})} d \mathbf{v}.
\label{eq:3.9}
 \end{equation}
Replacing equation (\ref{eq:3.8}) in the expression of the perturbed distribution, we then get

\begin{equation}
\tilde{W}_{\omega \mathbf{k}} =  - \frac{2 \pi Q}{M k^2} \frac{\delta (\omega - \mathbf{k} \cdot \mathbf{v}')}{\epsilon (\omega, \mathbf{k})} \frac{\mathbf{k} \cdot \partial W_0 / \partial \mathbf{v}}{(\omega - \mathbf{k} \cdot \mathbf{v})}.
\label{eq:3.10} 
\end{equation}
After inverse Fourier transforming, we can compute the perturbed force due to the test particle

\begin{equation}
\delta \mathbf{F} (\mathbf{r}, t) = - i \frac{2 Q \mathbf{k}}{k^2}  \int \frac{\exp [i \mathbf{k} \cdot (\mathbf{r} - \mathbf{r}_0 - \mathbf{v}' t)]}{\epsilon (\mathbf{k} \cdot \mathbf{v}', \mathbf{k})} \frac{d \mathbf{k}}{(2 \pi)^3}
\label{eq:3.11} 
\end{equation}
and the associated perturbation to distribution function

\begin{equation}
\tilde{W} (\mathbf{r}, \mathbf{v}, t) =  - \frac{Q}{M k^2} \int \frac{\exp [i \mathbf{k} \cdot (\mathbf{r} - \mathbf{r}_0 - \mathbf{v}' t)]}{\epsilon (\mathbf{k} \cdot \mathbf{v}', \mathbf{k})}\frac{\mathbf{k} \cdot \partial W_0 / \partial \mathbf{v}}{\mathbf{k} \cdot (\mathbf{v}' -  \mathbf{v})} \frac{d \mathbf{k}}{(2 \pi)^3}.
\label{eq:3.11b} 
\end{equation}
Finally, the density perturbation created at a position $\mathbf{r}$ and instant $t$ by a test particle located at position $\mathbf{r}_0$ at $t = 0$ moving with velocity $\mathbf{v}'$, can be now determined by

\begin{equation}
\tilde{n} (\mathbf{r},  t) =  \delta (\mathbf{r} - \mathbf{r}') + \int \tilde{W} (\mathbf{r}, \mathbf{v}, t) d \mathbf{v}.
\label{eq:3.12} 
\end{equation}
At this point, it should be notice the power of this method, as all the atoms in the gas can be individually considered as test particles. This means that the total averaged perturbations can be calculated by integrating the relevant quantities defined above over the equilibrium distribution function $W_0 (\mathbf{r}_0, \mathbf{v}')$. In particular, the averaged mean force perturbation can be determined by

\begin{equation}
< \delta \mathbf{F} (\mathbf{r},  t) > =  \int d \mathbf{r}_0 \int d \mathbf{v}' \delta \mathbf{F} (\mathbf{r}, t) W_0(\mathbf{v}). 
\label{eq:3.13} 
\end{equation}
This quantity is a linear superposition of purely oscillating quantities, and it can easily be found using equation (\ref{eq:3.11}) that it is identically zero

\begin{equation}
< \delta \mathbf{F} (\mathbf{r},  t) > \equiv 0,
\label{eq:3.13b} 
\end{equation}
while the averaged quadratic mean force $< | \delta \mathbf{F} (\mathbf{r},  t) |^2 >$, on the other hand, generally is not. It will be rather determined by
\begin{eqnarray}
< | \delta \mathbf{F} (\mathbf{r},  t) |^2 > =  \frac{Q^2}{V} \int d \mathbf{r}_0 \int d \mathbf{v}' \int \frac{d \mathbf{k}_1}{(2q \pi)^3} \frac{\mathbf{k}_1}{k_1^2} \frac{\exp [i \mathbf{k}_1 \cdot (\mathbf{r} - \mathbf{r}_0 - \mathbf{v}' t)]}{\epsilon (\mathbf{k}_1 \cdot \mathbf{v}', \mathbf{k}_1)} 
\\ \nonumber
\times \int \frac{d \mathbf{k}_2}{(2q \pi)^3} \frac{\mathbf{k}_2}{k_2^2} \frac{\exp [i \mathbf{k}_2 \cdot (\mathbf{r} - \mathbf{r}_0 - \mathbf{v}' t)]}{\epsilon (\mathbf{k}_2 \cdot \mathbf{v}', \mathbf{k}_2)}.
\label{eq:3.14} 
\end{eqnarray}
We next express the fluctuations inside the trap in terms of the statistical average over all test particles in the system.

\section{Density correlations}

Let us introduce $W_0 (\mathbf{v}) = n_0 f_0 (\mathbf{v})$, where the new distribution function $f_0 (\mathbf{v})$ is the normalized distribution function, such that

\begin{equation}
\int W_0 (\mathbf{v}) d \mathbf{v} = n_0 \int f_0 (\mathbf{v}) d \mathbf{v} = n_0.
\label{eq:4.1} 
\end{equation}
Using equations (\ref{eq:3.11b}) and (\ref{eq:3.12}), it is possible to express the atom density fluctuations due to a given test particle initially located at $\mathbf{r} (t = 0) = \mathbf{r}_0$ as

\begin{equation}
n (\mathbf{r}, t) = \delta (\mathbf{r} - \mathbf{r}_0 - \mathbf{v}' t) + \int \tilde{W} (\mathbf{r}, \mathbf{v}, t) d \mathbf{v},
\label{eq:4.2} \end{equation}
which can be more represented in a more suggestive fashion

\begin{equation}
n (\mathbf{r}, t) = \int \frac{d \mathbf{k}}{(2 \pi)^3} \left\{ 1 - \omega_p^2 \int \frac{\exp [i \mathbf{k} \cdot (\mathbf{r} - \mathbf{r}_0 - \mathbf{v}' t)]}{\epsilon (\mathbf{k} \cdot \mathbf{v}', \mathbf{k})} \frac{\mathbf{k} \cdot \partial W_0 / \partial \mathbf{v}}{\mathbf{k} \cdot (\mathbf{v}' -  \mathbf{v})} \right\}.
\label{eq:4.2b} 
\end{equation}
Here, we have introduced the definition of the equivalent plasma frequency for the neutral gas, $\omega_p^2 = Q n_0 / M$ \cite{mendkaiser}. Using the average as in Eq. (\ref{eq:3.13}), we can then state the density correlations as 

\begin{equation}
< n (\mathbf{r}_1, t_1) n (\mathbf{r}_2, t_2 )> = n_0 \int d \mathbf{r}_0 \int d \mathbf{v}' f_0 (\mathbf{v}') n (\mathbf{r}_1, t_1) n (\mathbf{r}_2, t_2).
\label{eq:4.3}
 \end{equation}
In explicit form, this can be written in the following way

\begin{equation}
< n (\mathbf{r}_1, t_1) n (\mathbf{r}_2, t_2 ) > = n_0 \int d \mathbf{r}_0 \int d \mathbf{v}' f_0 (\mathbf{v}') \int \frac{d \mathbf{k}_1}{(2 \pi)^3} e^{i \varphi_1} \int \frac{d \mathbf{k}_2}{(2 \pi)^3} e^{i \varphi_2} g (\mathbf{k}_1, \mathbf{v}') g (\mathbf{k}_2, \mathbf{v}'),
\label{eq:4.4}
 \end{equation}
where we have introduced the new auxiliary quantities
\begin{equation}
\varphi_j = i \mathbf{k}_j \cdot (\mathbf{r}_j - \mathbf{r}_0 - \mathbf{v}' t)
\label{eq:4.5} 
\end{equation}
and

\begin{equation}
g (\mathbf{k}_j, \mathbf{v}') = \left[ 1 - \omega_p^2 \int \frac{\mathbf{k}_j \cdot \partial f_0 / \partial \mathbf{v}}{k_j^2 \epsilon (\mathbf{k}_j \cdot \mathbf{v}', \mathbf{k}_j) \mathbf{k}_j \cdot (\mathbf{v}' - \mathbf{v})} \right].
\label{eq:4.6} 
\end{equation}
Integration over the initial position for the generic test particle $\mathbf{r}_0$ leads to the appearance of a delta function $\delta (\mathbf{k}_1 + \mathbf{k}_2)$ which, in its turn, allows us to integrate over $\mathbf{k}_2$. We then replace $\mathbf{k}_1$ by $\mathbf{k}$, in order to simplify the notation. By noting that the velocity $\mathbf{v}$ only appears in the parallel direction with respect to the wavemathbftor $\mathbf{k}$, we can write 

\begin{equation}
\mathbf{v} = u \frac{\mathbf{k}}{k} + \mathbf{v}_\perp
\label{eq:4.7}
\end{equation} 
and introducing the parallel distribution function

\begin{equation}
F_0 (u) = \int f_0 (\mathbf{v}) d  \mathbf{v}_\perp,
\label{eq:4.7b} 
\end{equation} 
we can finally write

\begin{equation}
< n (\mathbf{r}_1, t_1) n (\mathbf{r}_2, t_2 ) > = n_0 \int d u' F_0 (u') \int \frac{d \mathbf{k}}{(2 \pi)^3} e^{i \varphi} | I (k, u') |^2,
\label{eq:4.8}
 \end{equation}
where we have used the phase

\begin{equation}
\varphi = i \mathbf{k} \cdot (\mathbf{r}_1 - \mathbf{r}_2)  - i k u' (t_1 - t_2)
\label{eq:4.9} 
\end{equation}
and new argument

\begin{equation}
I (\mathbf{k}, u') = \left[ 1 + \omega_p^2 \int \frac{ \partial F_0 / \partial u}{k^2 \epsilon (k u', \mathbf{k}) (u' - u)} d u \right]
\label{eq:4.10}
 \end{equation}
 This is formally very similar to the results obtained for an electron-ion plasma \cite{book}, which is due to the Coulomb like interactions between nearby atoms in the ultra-cold neutral gas \cite{mendkaiser}. The main difference relies on the fact that we only have one particle species here, which suggests that the system can be regarded as a one-component plasma. We can now use these results to determine the structure factor discussed in Sec. II. Inserting Eqs. (\ref{eq:4.8}) to (\ref{eq:4.10}) in Eq.
(\ref{eq:2.11b}), and successively integrating over $\mathbf{r}_1$, $\mathbf{r}_2$, $t_1$, $t_2$, $\mathbf{k}$, and $u'$, we finally obtain 

\begin{equation}
S(k',\omega') = \frac{F_0 (\omega'/k')}{k'}  | I (k' ,\omega') |^2,
\label{eq:4.11}
\end{equation}

where we have used
\begin{equation}
I (k', \omega') = \left[ 1 + \omega_p^2 \int \frac{ (\partial F_0 / \partial u) d u}{k^{'2} \epsilon (k', \omega') (u - \omega'/k')}  \right].
\label{eq:4.12} 
\end{equation}
In order to illustrate the collective behavior of the system during the scattering, we assume that the atoms in the trapped approximately follow a Maxwell-Boltzmann distribution

\begin{equation}
F_{0}(v')=\frac{1}{\sqrt{\pi}v_{th}}e^{-v'^2/v_{th}^2},
\label{eq:4.13}
\end{equation} 
where $v_{th}=\sqrt{2k_{B}T/m}$ represents the thermal velocity of the atoms, $k_{B}$ is the Boltzmann constant and $T$ is the temperature. Defining the scattering parameter $\alpha=k'\lambda_{D}$, where $\lambda_{D}=v_{th}/\omega_{P}$ is the Debye length, the dynamical structure factor can be explicitly given by

\begin{equation}
S(\alpha,\omega')=\frac{e^{-\alpha\omega'/\omega_{p}}}{\sqrt{\pi}\alpha}I(\alpha,\omega'/\omega_{p}),
\label{eq:4.14}
\end{equation} 
where $\alpha=k'\lambda_{D}$ represents the scattering parameter. The latter defines the nature of the scattering processes. Moreover, for $\alpha \lesssim 1$, the scattering signal results from a  coherent process, as the system is essentially composed of interacting particles. In a plasma language, it means that the Debye length $\lambda_{D}$ is larger or comparable to the perturbation wavelength $2\pi/k'$ and therefore the correlations between the atoms play an important role on the scattering signal (unscreened, correlated atoms); on the contrary, if $\alpha \gg 1$, the scattering is said to be incoherent and the Debye length is assumed to be much smaller than the typical size of the perturbation (screened, uncorrelated atoms). In the figure, we have ploted the normalized structure factor in Eq. (\ref{eq:4.14}) for three different values of $\alpha$. We observe a strong resonance near $\omega'=\omega_{p}$ for the case of coherent scattering ($\alpha=0.85$ and $\alpha=1.0$), while a very broad spectrum is obtained for the case of incoherent scattering ($\alpha=5.0$). The physical reason for such results are related with the interference mechanism during the scattering: the excitation of collective perturbations of the density will result on the constructive interference (and therefore amplification) of the scattered signal around the natural frequency characterizing the long-range order of the interaction, i.e. $\omega_{p}$. On the contrary, if uncorrelated, single-atoms fluctuations are excited in the systems, the interference will be destructive and, therefore, no resonance is expected to be observed in the spectrum $S(\alpha,\omega')$.\par

Few words are now devoted to discuss the experimental observation of the spectra. The scattering mechanism presented in this work is similar to Rayleigh scattering, with the important difference that long-range order between the atoms is present, which is naturally induced during the laser cooling process. Therefore, this could be experimentally implemented in magneto-optical traps in order to measure the collective excitations of the system, as both coherent and incoherent regimes of scattering are possible, opening the door to a new diagnosis. A simple setup for the present technique would consist of a external probe laser beam with a very small waist (much smaller than those of the cooling beams) propagating inside the cloud. The resulting fluorescent light, modulated by the density fluctuations, can be focused into a photodetector placed at a certain angle $\theta_{0}$ with respect to the probe beam direction. After Fourier transform the photodetector signal, both the scattering parameter $\alpha$ and the effective plasma frequency $\omega_{p}$ can be measured by fitting the fluorescence signal with Eq. (\ref{eq:4.14}). In order to distinguish the scattering light from the fluorescence signal coming from the MOT, the probe beam should operate a large detuning compared to the cooling detuning, i.e ($\vert \delta_{0}\vert \gg \vert \delta_{cooling}\vert$).   

\begin{figure}
\includegraphics{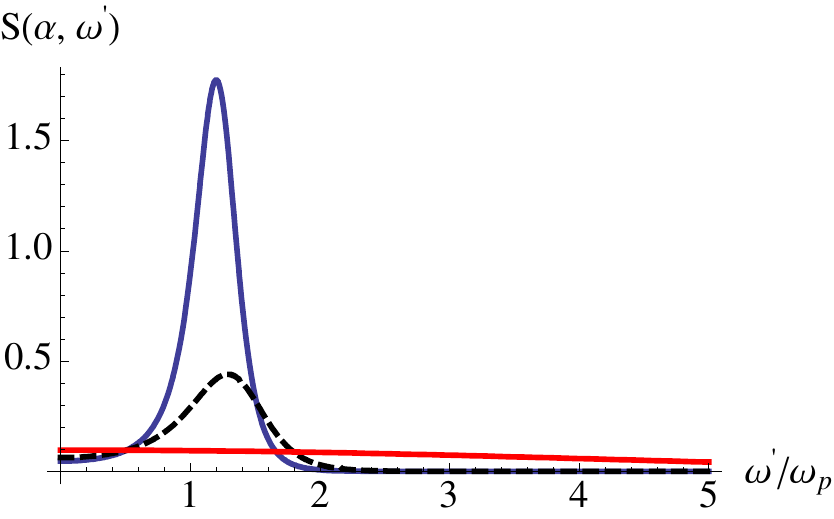}
\label{scatt}
\caption{Dynamical structure factor $S(\alpha,\omega')$ for different values of the scattering parameter $\alpha$: blue full line ($\alpha=0.85$), black dashed line ($\alpha=1.0$), and red full line ($\alpha=5.0$). Both coherent and incoherent scattering processes are represented. }
\end{figure}

\section{Conclusions}

In this work we have determined the spectrum of density correlations of ultra-cold atoms confined in a magneto-optical trap. We have explored the formal analogies between the physics of the neutral atom gas at that of a plasma medium. The existence of such analogies is related with the repulsive force between two nearby atoms in the ultra-cold gas, associated with the exchange of scattered photons from the laser cooling system. This repulsive interaction can be described  by Poisson like equations for the collective force, similar to that describing electrostatic fields in a real plasma medium.

Here we have adapted the test particle approach, well known in plasma physics, to the case of the neutral cold gas. We have determined the spectrum of density correlations in the gas and we have established the corresponding spectra of laser light scattering. We have applied this to a medium in thermal equilibrium and given examples of expected spectra. This work proposes the use of laser light scattering as a diagnostic technique for probing the internal kinetic properties of a ultra-cold gas, and could be useful for the interpretation of future experiments in this field.

\end{document}